\documentclass[12pt]{article}
\usepackage{cite}
\usepackage{enumerate}
\usepackage{ifpdf}
\usepackage{ae}
\usepackage[T1]{fontenc}
\usepackage[ansinew]{inputenc}
\usepackage{amsmath}
\usepackage{relsize}
\usepackage{amssymb}
\usepackage{tikz}
\usepackage{graphicx}
\usepackage{color}
\definecolor{darkblue}{cmyk}{0.9,0.9,0,0}
\definecolor{darkgreen}{rgb}{0,0.55,0}
\usepackage[colorlinks=true,linkcolor=darkblue,citecolor=darkblue,urlcolor=darkblue]{hyperref}
\usepackage{epsfig}
\usepackage{graphicx}

\newcommand{\beq}{\begin{equation}}
\newcommand{\eeq}{\end{equation}}
\newcommand{\beqy} {\begin{eqnarray}}
\newcommand{\eeqy} {\end{eqnarray}}
\newcommand{\bsmat}{\begin{smallmatrix}}
\newcommand{\esmat}{\end{smallmatrix}}
\newcommand{\bmat}{\begin{matrix}}
\newcommand{\emat}{\end{matrix}}
\newcommand{\sfrac}[2]{{\textstyle\frac{#1}{#2}}}

\def\({\left(}
\def\){\right)}
\def\[{\left[}
\def\]{\right]}

\def\<{\langle}
\def\>{\rangle}

\usepackage{varioref}
\usepackage{makeidx}
\makeindex

\usepackage[english]{babel}
\usepackage{subfig}

\usepackage{tabularx}
\begin{document}

\thispagestyle{empty}

\renewcommand{\thefootnote}{\fnsymbol{footnote}}
\setcounter{page}{1}
\setcounter{footnote}{0}
\setcounter{figure}{0}

\begin{titlepage}

\begin{center}

\vskip 2.3 cm 

\vskip 5mm

{\Large \bf Lessons from crossing symmetry at large $N$ }
\vskip 0.5cm

\vskip 15mm

{Luis F. Alday, Agnese Bissi and Tomasz {\L}ukowski}

\vspace{1cm}
\centerline{{\it Mathematical Institute, University of Oxford,}}
\centerline{{\it Andrew Wiles Building, Radcliffe Observatory Quarter,}}
\centerline{{\it Woodstock Road, Oxford, OX2 6GG, UK}}

\end{center}

\vskip 2 cm

\begin{abstract}
\noindent  We consider the four-point correlator of the stress tensor multiplet in ${\cal N}=4$ SYM. We construct all solutions consistent with crossing symmetry in the limit of large central charge $c \sim N^2$ and large $g^2 N$. While we find an infinite tower of solutions, we argue most of them are suppressed by an extra scale $\Delta_{gap}$ and are consistent with the upper bounds for the scaling dimension of unprotected operators observed in the numerical superconformal bootstrap at large central charge. These solutions organize as a double expansion in $1/c$ and $1/\Delta_{gap}$. Our solutions are valid to leading order in $1/c$ and to all orders in $1/\Delta_{gap}$ and reproduce, in particular, instanton corrections previously found.
Furthermore, we find a connection between such upper bounds and positivity constraints arising from causality in flat space. Finally, we show that certain relations derived from causality constraints for scattering in $AdS$ follow from crossing symmetry. 
\end{abstract}

\end{titlepage}

\setcounter{page}{1}
\renewcommand{\thefootnote}{\arabic{footnote}}
\setcounter{footnote}{0}

\section{Introduction}
Conformal field theories (CFT) are one of the pillars of theoretical physics. Important motivations to study them are their role in phase transitions and their relation to renormalization group flows. Over the last two decades, it also became evident that they describe quantum gravity in $AdS$ space, through the $AdS/CFT$ correspondence. The main ingredient of a CFT is the set of local primary operators ${\phi}(x)$ and the main observables are correlators of these operators
\begin{equation}
\langle {\phi}_1(x_1) \cdots  {\phi}_n(x_n) \rangle\,.
\end{equation}
A CFT is defined by its {\it CFT data}, namely, the list of scaling dimensions $\Delta_i$ of all local primary operators and the operator product expansion (OPE) coefficients $c_{ijk}$ for any three primaries. In a unitary CFT this data satisfies certain constraints. In particular the OPE coefficients are real numbers and for the case to be studied in this paper
\begin{equation}
\Delta \geq \ell+2\,,
\end{equation}
for a primary operator of spin $\ell$. Once the CFT data is given, the OPE allows to write, in principle, {\it any} higher point correlator. The idea of the conformal bootstrap program is to use crossing symmetry of correlation functions, together with unitarity and the structure of the OPE, in order to constrain the CFT data. In the simplest setting, which is also the relevant for this paper, we consider the correlator of four identical operators of scaling dimension $\Delta$. Conformal symmetry implies
\begin{equation}
\label{4pt}
\langle {\phi}(x_1) {\phi}(x_2){\phi}(x_3) {\phi}(x_4) \rangle = \frac{{g}(u,v)}{x_{12}^{2\Delta}x_{34}^{2\Delta}}\,,
\end{equation}
where we have introduced the cross-ratios
\begin{equation}
u=\frac{x_{12}^2 x_{34}^2}{x_{13}^2 x_{24}^2},~~~v=\frac{x_{14}^2 x_{23}^2}{x_{13}^2 x_{24}^2}\,.
\end{equation}
By using the OPE we can decompose the correlator \eqref{4pt} as a sum over intermediate states ${\cal \phi}_i$ along the $(12)(34)$ channel
\begin{equation}
g(u,v) =1+ \sum_{{\cal \phi}_i} c_{i}^2 G_{i}(u,v).
\end{equation}
The sum runs over the conformal primary intermediate states appearing in the OPE of $\phi \times \phi$. The conformal blocks $G_{i}(u,v)$ are completely fixed by conformal symmetry and depend only on the dimension and spin of the intermediate state. They encode the contribution of a given primary together with its tower of descendants. In the above expansion we have also singled out the contribution from the identity operator, always present in the OPE of two identical operators. We could have instead chosen to expand along the $(13)(24)$ channel, and the result should have been the same. Indeed, crossing symmetry of the four-point function implies
\begin{equation}
v^\Delta g(u,v) = u^\Delta g(v,u),
\end{equation}
which results in the following non-trivial equation involving the CFT data
\begin{equation}
\label{usualcb}
\sum_i c_{i}^2 \left(v^\Delta G_{i}(u,v) - u^\Delta G_{i}(v,u) \right) = u^\Delta - v^\Delta\,.
\end{equation}
Note that the r.h.s.~arose from the presence of the identity operator. Equation \eqref{usualcb} is called the conformal bootstrap equation. 

So far the discussion has been pretty general. However, specific conformal field theories often possess extra symmetries which impose extra constraints. An important example is that of supersymmetric conformal field theories (SCFT). The subject of this paper will be four-dimensional ${\cal N}=4$ super Yang-Mills (SYM). This is the maximally symmetric four dimensional conformal field theory, and is particularly interesting since it describes quantum gravity on $AdS$ space. 

In ${\cal N}=4$ super Yang-Mills the energy-momentum tensor lies in a half-BPS multiplet, whose superconformal primary is a scalar operator ${\cal O}$ of protected dimension two and which transforms in the ${\bf 20'}$ representation of the $SU(4)$ R-symmetry group\footnote{In order to simplify the notation we will obviate the representation index.}. In \cite{Beem:2013qxa}, the consequences of crossing-symmetry of the correlator $\langle {\cal O} {\cal O}{\cal O}{\cal O} \rangle$ were analyzed and were written in the form of a (super)conformal bootstrap equation:
\begin{equation}
\label{cbo}
\sum_{\Delta,\ell}a_{\Delta,\ell} \left( G_{\Delta,\ell}(u,v) -G_{\Delta,\ell}(v,u) \right)= F_{short}(u,v,c).
\end{equation}
Although the derivation is conceptually similar to the previous case, there are important differences. First, among the states in the OPE of ${\cal O} \times {\cal O}$ there is a rich spectrum of protected operators, belonging to short or semi-short multiplets, which do not acquire anomalous dimension and whose OPE coefficient is fixed due to superconformal Ward identities \cite{Nirschl:2004pa}.  The r.h.s.~of \eqref{cbo} resums the contribution from all such operators, instead of just the identity. The structure of $F_{short}(u,v,c)$ is very rich, but it is important to note that it is only a function of the central charge, and not of the coupling constant of the theory, and is explicitly known. Second, supersymmetry relates operators in different  conformal towers. Therefore, the sum runs only over unprotected superconformal primaries in long multiplets and correspondingly $G_{\Delta,\ell}(u,v)$ are the superconformal blocks \cite{Nirschl:2004pa}, whose explicit expression is given below. Finally, $a_{\Delta,\ell}$ denote the square of the OPE coefficients and are non-negative as a consequence of unitarity.

In spite of fitting in one line both \eqref{usualcb} and \eqref{cbo} are formidable equations: they involve a double infinite sum, over an unknown spectrum with unknown OPE coefficients. But whatever these unknowns are, they should combine (at each value of the coupling constant!) to give the explicitly known right hand side. In \cite{Rattazzi:2008pe} it was understood how to efficiently use these kind of equations. The idea is to propose a putative spectrum. For a given spectrum, the r.h.s.~of \eqref{cbo} will be a linear combination in terms of the basis of functions  $\left( G_{\Delta,\ell}(u,v) -G_{\Delta,\ell}(v,u) \right)$. If either of the coefficients $a_{\Delta,\ell}$ turns out to be negative, then the putative spectrum is not consistent with unitarity and can be ruled out. In practice this is done numerically, and only the support of the putative spectrum for each $\ell$ needs to be specified. The positivity of $a_{\Delta,\ell}$ is checked by acting on both sides of \eqref{cbo} with families of linear operators. For a given spin $\ell$, we define the twist as $\tau=\Delta-\ell$. The leading twist operator for a given spin is the operator with the smallest twist. The method above gives numeric upper bounds for the dimension of leading twist operators, see \cite{Rattazzi:2008pe} for the details. The authors of \cite{Beem:2013qxa} carried out this analysis and found numerical bounds for the dimension of unprotected leading twist operators of low spin $\ell=0,2,4$, for various values of the central charge $c=(N^2-1)/4$. At large values of the central charge, the upper bounds found by  \cite{Beem:2013qxa} were consistent with 
\begin{eqnarray}
\Delta_0 &=& 4 - \frac{16}{N^2}+\ldots\,,\\
\Delta_2 &=& 6 - \frac{4}{N^2}+\ldots\,,\\
\Delta_4 &=& 8 - \frac{48}{25 N^2}+\ldots\,.
\end{eqnarray}
These results are precisely the dimensions $\Delta_\ell$ at large $N$ found from a holographic supergravity computation \cite{D'Hoker:1999jp,Arutyunov:2000ku,Dolan:2001tt}!

The aim of this paper is to construct analytic solutions to the conformal bootstrap equation \eqref{cbo} at large values of the central charge. Solutions at large $N$ consistent with crossing symmetry for non-supersymmetric four dimensional conformal field theories have already been constructed in \cite{Heemskerk:2009pn}\footnote{Analytic studies of the crossing relations also include \cite{Fitzpatrick:2012yx,Komargodski:2012ek,Alday:2013cwa} but in these works the focus is on operators with small twist but very high spin.}. We start section \ref{Secion:solutions} by applying their methods to our case. In addition to a large central charge, we assume single trace operators acquire a parametrically large dimension, which happens for large $\lambda=g^2N$. As in \cite{Heemskerk:2009pn}, an infinite number of solutions is found. We then consider these solutions in Mellin space, and show that they take a remarkably simple form. The Mellin expressions allow to understand several features of these solutions, which are useful for their interpretation carried out in section \ref{Secion:Interpretation}. Unitarity requires the introduction of a gap scale $\Delta_{gap}$ at which new operators enter into the spectrum. We argue that the infinite tower of solutions is suppressed by powers of $\Delta_{gap}$ and the full solution takes the form of a double expansion. This is closely reminiscent of the picture of  \cite{Fitzpatrick:2010zm}. Comparing our results with instanton contributions to the four point function, previously computed in the literature, justifies this picture and sets $\Delta_{gap} \sim N^{1/4}$ for intermediate values of the coupling constant. At the end of section \ref{Secion:Interpretation} we show that even if the first two "extra" solutions are not suppressed, positivity constraints from causality in flat space would still imply consistency with the upper bounds found by  \cite{Beem:2013qxa}. We find this connection quite remarkable. Furthermore, we show that certain relations derived from causality constraints for scattering in $AdS$ found in \cite{malda} follow from crossing symmetry. We end up with some conclusions. 


\section{Analytic solutions at large \texorpdfstring{$N$}{}}
\label{Secion:solutions}
\subsection{The superconformal bootstrap equation}
As already mentioned in the introduction, the conformal bootstrap equations for ${\cal N}=4$ SYM takes the form
\begin{equation}
\label{cb}
\sum_{\Delta,\ell}a_{\Delta,\ell} F_{\Delta,\ell}(u,v)= F_{short}(u,v,c),
\end{equation}
where the sum runs over unprotected superconformal primaries, in the singlet representation of the $R-$symmetry group $SU(4)$, with spin $\ell=0,2,4,\ldots$, and with spectrum satisfying the unitarity bound $\Delta \geq \ell+2$. 
For brevity, we have introduced
\begin{equation}
F_{\Delta,\ell}(u,v)=G_{\Delta,\ell}(u,v) - G_{\Delta,\ell}(v,u)\,.
\end{equation}
In order to write down the explicit expressions for the superconformal blocks it is convenient to introduce variables $z,\bar z$, with $u = |z|^2,~v = |1-z|^2$. In terms of these\footnote{We will use both sets, $(u,v)$ and $(z,\bar z)$, interchangeably.}
\begin{equation}
\label{sublock}
G_{\Delta,l}(z,\bar z)=(1-z)^2(1-\bar z)^2(z\bar z)^{\frac{1}{2}(\Delta-l)}g_{\Delta+4,l}(z,\bar z)
\end{equation}
with
\begin{equation}
g_{\Delta,l}(z,\bar z)=\left(\frac{-1}{2}\right)^l \frac{1}{z-\bar z}\left(z^{l+1}k_{\Delta+l}(z)k_{\Delta-l-2}(\bar z)-(\bar z)^{l+1}k_{\Delta+l}(\bar z)k_{\Delta-l-2}(z)\right)
\end{equation}
and 
\begin{equation}
k_{\beta}(z)={}_2F_{1}\left(\frac{\beta}{2},\frac{\beta}{2},\beta,z\right).
\end{equation}
Alternatively, we can also write
\begin{equation}
\sum_{\Delta,\ell}a_{\Delta,\ell} G_{\Delta,\ell}(u,v)=  G_{short}(u,v,c) + {\cal A}(u,v)\,,
\end{equation}
where the function $G_{short}(u,v,c)$ is related to $F_{short}(u,v,c)$ by
\begin{equation}
 F_{short}(u,v,c) = G_{short}(u,v,c)  -  G_{short}(v,u,c) \,
\end{equation}
and its explicit form is given in appendix \ref{App:Gshort}.
Then the conformal bootstrap equation \eqref{cb} is equivalent to the requirement
\begin{eqnarray}
\label{crossingA}
 {\cal A}(u,v) = {\cal A}(v,u)\,.
\end{eqnarray}
Let us emphasize that ${\cal A}(u,v)$ generally depends on the coupling constant and in order to compute it one usually has to resort to explicit computations. The superconformal bootstrap equation differs from the standard one in two aspects:
\begin{itemize}
\item It involves superconformal blocks, instead of conformal blocks. For the present case they are proportional to the usual conformal blocks upon a shift $\Delta \rightarrow \Delta+4$. 
\item $F_{short}(u,v,c)$ has a much richer structure than its analogue in conformal field theories, which usually contains only the identity operator. 
\end{itemize}


\subsection{Solutions at large \texorpdfstring{$N$}{}}
\label{analyticsolutionslarge}
We look for solutions consistent with crossing symmetry in the large $N$ expansion up to order $1/N^2$. We will assume that single trace operators acquire a parametrically large dimension. More precisely, in addition to $N$ large we are assuming $\lambda=g^2 N$ is also large. Hence the space of intermediate states is spanned by double trace operators of the form ${\cal O}_{n,\ell}={\cal O} \partial_{\mu_1}\ldots \partial_{\mu_\ell}(\partial . \partial)^n {\cal O}$, labelled by integers $n=0,1,\ldots$ and $\ell=0,2,\ldots$, of dimension $2n+\ell+4$ at leading order, and spin $\ell$. Higher trace operators will not contribute to the order we are working at. The function $F_{short}(u,v,c)$ has a very simple expansion in $1/N^2$ or rather the inverse of the central charge $c$
 \begin{equation}
 F_{short}(u,v,c) =  F^{(0)}_{short}(u,v) + \frac{1}{c}F^{(1)}_{short}(u,v) \,.
 \end{equation}
At leading order the conformal bootstrap equation reduces to
\begin{equation} \sum_{n=0}^\infty
\sum_{\substack{\ell=0\\ \textrm{even}}}^\infty a^{(0)}_{n,\ell} F_{4+2n+\ell,\ell}(u,v)= F^{(0)}_{short}(u,v)\,.
\end{equation}
The set of function $F_{4+2n+\ell,\ell}(u,v)$ is a complete set, hence given $F^{(0)}_{short}(u,v)$ we can fix the structure constants at leading order. We obtain
\begin{equation}
a^{(0)}_{n,\ell}= \frac{2^{-7-\ell-4 n} \pi (1+\ell) (6+\ell+2 n)   \Gamma(3+n) \Gamma(4+\ell+n)}{\Gamma(\frac{5}{2}+n)\Gamma(\frac{7}{2}+\ell+n)}\,.
\end{equation}
In order to find solutions at the next order, we expand the ingredients of the conformal bootstrap equation as follows
\begin{eqnarray}
\Delta_{n,\ell} &=& 4+2n+\ell+ \frac{1}{N^2} \gamma_{n,\ell}+\ldots\,,\\
 a_{n,\ell}  &=& a^{(0)}_{n,\ell} +\frac{1}{N^2} a^{(1)}_{n,\ell} +\ldots\,,\\
G_{short}(u,v,c) &=& G^{(0)}_{short}(u,v)+\frac{1}{N^2}G^{(1)}_{short}(u,v)+\ldots\,, \\
 {\cal A}(u,v) &=& \frac{1}{N^2} A(u,v)+\ldots \,.
\end{eqnarray}
Then at order $1/N^2$ we obtain
\begin{eqnarray}
\label{firstorder}
\sum_{n=0}^\infty\sum_{\substack{\ell=0\\ \textrm{even}}}^\infty  \left( a^{(1)}_{n,\ell}  G_{4+2n+\ell,\ell}(z,\bar z) + a^{(0)}_{n,\ell}  \gamma_{n,\ell} \frac{1}{2} \frac{\partial}{\partial n} G_{4+2n+\ell,\ell}(z,\bar z) \right)  = G^{(1)}_{short}(z,\bar z)+A(z,\bar z).
\end{eqnarray}
We need to find sets $\{  \gamma_{n,\ell}, a^{(1)}_{n,\ell}\}$ which lead to a r.h.s.~consistent with crossing symmetry \eqref{crossingA}, namely $A(z,\bar z)= A(1-z,1-\bar{z})$. It is easy to see that $A(z,\bar z)$ has to be different from zero. Indeed, on the l.h.s.~of \eqref{firstorder} the operators of leading twist $\tau=\Delta - \ell$, have twist four, which corresponds to $n=0$, so an expansion of the l.h.s.~in powers of $u$ will start with $u^2$ \footnote{In general, the small $u$ behavior of the conformal block for an operator of twist $\tau$ is $u^{\tau/2}$.}. On the other hand $G^{(1)}_{short}(u,v)$ has the small $u$ expansion
\begin{equation}
G^{(1)}_{short}(u,v) = 16\, u\, v \,\frac{1-v^2+2v \log v}{(1-v)^3}+\ldots\,.
\end{equation} 
Therefore  $A(z,\bar z)$ has to cancel that contribution. The minimal choice is that given by the supergravity result \cite{Dolan:2006ec}
\begin{eqnarray}
A(z,\bar z)= -16\, u^2 v^2 \bar D_{2422}(z,\bar z),
\end{eqnarray}
where the definition of the $\bar D$ functions is given in the appendix \ref{App:Dbar}. Performing the conformal partial wave expansion, this leads to specific values for $\{  \gamma_{n,\ell}, a^{(1)}_{n,\ell}\}$, which we denote  $\{  \gamma^{sugra}_{n,\ell}, a^{(1),sugra}_{n,\ell}\}$. In particular one obtains
\begin{eqnarray}
\label{sugraresult}
\gamma^{sugra}_{n,\ell}&=&-\frac{4 (1+n) (2+n) (3+n) (4+n)}{(1+\ell) (6+\ell+2 n)} \,,\\
a^{(1),sugra}_{n,\ell} &=& \frac{1}{2}\frac{\partial}{\partial n} \left( a^{(0)}_{n,\ell}\gamma^{sugra}_{n,\ell} \right)\,.
\end{eqnarray}
Now the general solution to \eqref{firstorder} can be written as
\begin{eqnarray}
\gamma_{n,\ell}&=&\gamma^{sugra}_{n,\ell} + \hat \gamma_{n,\ell}\,,\\
a^{(1)}_{n,\ell} &=& a^{(1),sugra}_{n,\ell} + \hat a^{(1)}_{n,\ell} \,,
\end{eqnarray}
where $\{  \hat \gamma_{n,\ell}, \hat a^{(1)}_{n,\ell}\}$ are solutions of the "homogeneous" equation
\begin{gather}
\label{firstorderhom}
\sum_{n=0}^\infty\sum_{\substack{\ell=0\\ \textrm{even}}}^\infty \left( \hat a^{(1)}_{n,\ell} G_{4+2n+\ell,\ell}(z,\bar z) + a^{(0)}_{n,\ell}  \,\hat \gamma_{n,\ell}\, \frac{1}{2} \frac{\partial}{\partial n} G_{4+2n+\ell,\ell}(z,\bar z) \right)  = A(z,\bar z), \nonumber \\
A(z,\bar z)=A(1-z,1-\bar z)
\end{gather}
In order to construct explicit solutions to (\ref{firstorderhom}) we follow closely \cite{Heemskerk:2009pn} adapted to our case. The idea is to restrict ourselves to solutions with intermediate operators of spin up to a maximum value $L$, namely we allow $\ell=0,2,\ldots,L$. Next, we consider the limit $z \rightarrow 0$ and $\bar z \to 1$ and focus in the terms proportional to $\log z \log (1-\bar z)$. This isolates the contributions from the anomalous dimensions and we obtain the following set of conditions for any pair  $(p,q)\in\mathbb{Z}_+^2$
\begin{align}\nonumber
0= \sum_{n=0}^\infty\sum_{\substack{\ell=0\\ \textrm{even}}}^\infty \frac{1}{2^\ell}a^{(0)}_{n,\ell}  \hat \gamma_{n,\ell} \left\{I(n+3,q)\delta_{\ell+n+2,p-1}-I(n+\ell+4,q)\delta_{n+1,p-1}\right\}\\ \label{gammaeq}
+ \sum_{n=0}^\infty\sum_{\substack{\ell=0\\ \textrm{even}}}^\infty \frac{1}{2^\ell}a^{(0)}_{n,\ell} \hat \gamma_{n,\ell} \left\{I(\ell+n+4,p)\delta_{n+1,q-1}-I(n+3,p)\delta_{\ell+n+2,q-1}\right\}\,,
\end{align}
where have defined
\begin{equation}
I(m,m')=\oint \frac{dz}{2\pi i}\frac{(1-z)^{m-3}}{z^{m'-1}}\tilde k_{2m}(z)k_{-2m'}(z),
\end{equation}
with
\begin{equation}
\tilde k_{\beta}(z)=-\frac{\Gamma(\beta)}{\Gamma^2(\beta/2)}{}_2 F_{1}\left(\frac{\beta}{2},\frac{\beta}{2},1,z\right).
\end{equation}
The counting of solutions of \eqref{gammaeq} works exactly as in \cite{Heemskerk:2009pn}. For instance, for $L=0$ there is exactly one solution, proportional to an overall normalization factor, for which
\begin{equation}
\hat \gamma_{n,0} =  - \frac{ (1+n)^2 (2+n)^2 (3+n)^3 (4+n)^2 (5+n)^2}{960 (5+2 n) (7+2 n)}\,.
\end{equation}
 For $L=2$ there are two new solutions, for $L=4$ there are three new solutions, and so on. For a cut-off $L$ the total number of solutions is $(L+2)(L+4)/8$. After having found $\hat \gamma_{n,\ell}$, we look at the terms proportional to $\log(1-\bar z)$ in \eqref{firstorderhom}. In all the cases we find
\begin{equation}
 \hat a^{(1)}_{n,\ell} = \frac{1}{2}\frac{\partial}{\partial n} \left( a^{(0)}_{n,\ell}\hat \gamma_{n,\ell} \right),
\end{equation}
exactly as for the supergravity solution. This was also observed for the solutions in \cite{Heemskerk:2009pn} and was subsequently proven by \cite{Fitzpatrick:2011dm}. To each solution corresponds a  function $A(z,\bar z)$. We denote by $A^{(L)}_m(z,\bar z)$ for $m=0,2,4,\ldots,L$, the new solutions that appear at spin $L$ (later we will be more specific about the index $m$). In the next subsection we will show that these solutions admit a simple representation in Mellin space, and we will give an analytic expression for all of them, but in the meantime let us add that each of these solutions can be written in terms of $\bar D$ functions, and so have an interpretation in terms of Witten diagrams, as expected. For instance
\begin{eqnarray}
A^{(0)}_0(u,v) &=& u^2 v^2 \bar D_{4444}(u,v) ,\\
A^{(2)}_0(u,v)  &=& u^2 v^2(1+u +v)\bar D_{5555}(u,v),\\
 A^{(2)}_2(u,v)  &=&u^2 v^2\left(\bar D_{5656}(u,v)+\bar D_{6565}(u,v) + u^2 \bar D_{6655}(u,v)+\right. \\
\nonumber & &\left.+ u \bar D_{5566}(u,v)+ v^2 \bar D_{5665}(u,v)+ v \bar D_{6556}(u,v)\right).
\end{eqnarray}
Note that these expressions agree with the ones found holographically by \cite{Heemskerk:2009pn}, for the special case $\Delta=4$, however, our external states have dimension two. 


\subsection{Solutions in Mellin space}
As shown in \cite{Penedones:2010ue,Fitzpatrick:2011ia, Paulos:2011ie}, beautiful structure emerges when expressing correlators in Mellin space, specially in the large $N$ limit.  For the purposes of the present note, given a function of cross ratios $A(u,v)$, we define its Mellin representation $M(x,y)$ by
\begin{equation}\label{Mellin}
A(u,v) =\frac{1}{(2\pi i)^2} \int  \Gamma^2(x)\Gamma^2(y) \Gamma^2(2-x-y)M(x,y) u^{-x} v^{-y} dx dy\,,
\end{equation}
where the integration contours are over the imaginary axis shifted by a small positive real part. Notice that we defined \eqref{Mellin} in such a way that the Mellin amplitude for $\bar D_{2222}(u,v)$ equals 1. The solutions $A^{(L)}(u,v)$ we have found in the previous section\footnote{We denote by $A^{(L)}(u,v)$ the collective space of solutions entering at spin $L$.} possess two important symmetries. First, due to crossing symmetry they satisfy
\begin{equation}
A^{(L)}(u,v)=A^{(L)}(v,u).
\end{equation}
Second, they are obtained from a conformal partial wave decomposition in the $(12)(34)$ channel, as such
\begin{equation}
A^{(L)}(u,v)= v^2 A^{(L)}\left(\frac{u}{v},\frac{1}{v}\right).
\end{equation}
This is a symmetry of each conformal block \eqref{sublock} separately, and physically corresponds to exchanging operators 1 and 2. These conditions will translate as symmetries in Mellin space 
\begin{eqnarray}
\label{sym}
M^{(L)}(x,y)&=&M^{(L)}(y,x),\\ \label{sym2}
M^{(L)}(x,y)&=&\frac{\Gamma^2(-2-x-y)\Gamma^2(4+y)}{\Gamma^2(y)\Gamma^2(2-x-y)} M^{(L)}(x,-2-x-y).
\end{eqnarray}
For instance, we can work out the solution of previous section for $L=0$ in Mellin space, we obtain
\begin{equation}
\label{m0}
M^{(0)}(x,y)=  \frac{x^2 (1+x)^2 y^2 (1+y)^2}{(1-x-y)^2 (x+y)^2}\,,
\end{equation}
which can be easily checked to satisfy both symmetries. 

In order to construct the solutions to crossing symmetry in Mellin space for higher spin $L$, we work out the Mellin representation for the superconformal blocks \eqref{sublock}. This is given by
\begin{equation}
{\cal B}_{\Delta,\ell}(x,y)= \frac{y^2 (1+y)^2}{(-1+x+y)^2 (x+y)^2} B_{\Delta+4,\ell}(x,y),
\end{equation}
where $B_{\Delta,\ell}(x,y)$ are the usual four dimensional conformal blocks in Mellin space, for the exchange of a particle of dimension $\Delta$ and spin $\ell$ and for external particles of dimension two. They have been constructed in  \cite{Mack:2009mi,Fitzpatrick:2011hu} and are given by
\begin{equation}
B_{\Delta,\ell}(x,y) =\frac{e^{-i \pi  \Delta } \left(e^{i \pi  (\ell-2 x+\Delta )}-1\right) \Gamma(-\frac{\ell}{2}+x-\frac{\Delta }{2}) \Gamma(-2-\frac{\ell}{2}+x+\frac{\Delta }{2})}{\Gamma^2(x)} P^{(\ell)}_\Delta(x,y),
\end{equation}
where $P^{(\ell)}_\Delta(x,y)$ is a polynomial of degree $\ell$, defined in \cite{Mack:2009mi}, and whose explicit form will not be important for us. Note the remarkable fact that the dependence of ${\cal B}_{\Delta,l}(x,y)$ on $y$ is very simple. This is a very nice feature of Mellin space and it will be important in the construction of our solutions. 
Let us see how $M^{(0)}(x,y)$ follows directly from symmetries \eqref{sym} and \eqref{sym2} and the expression for conformal blocks in Mellin space. Since this solution involves only intermediate states with $\ell=0$, it is a sum of terms ${\cal B}_{\Delta,0}(x,y)$. Therefore,  its $y$ dependence is fixed, and it should take the form
\begin{equation}
M^{(0)}(x,y)= \frac{y^2 (1+y)^2}{(-1+x+y)^2 (x+y)^2} f(x).
\end{equation}
But then the symmetries \eqref{sym} and \eqref{sym2} fix $f(x) = x^2(1+x)^2$ up to a constant! Hence we reobtain \eqref{m0}. This reasoning can be extended to higher spins. Allowing intermediate states up to spin $L$ we obtain
\begin{equation}\label{masp}
M^{(L)}(x,y)= \frac{x^2 (1+x)^2 y^2 (1+y)^2}{(1-x-y)^2 (x+y)^2} P^{(L)}(x,y),
\end{equation}
where $P^{(L)}(x,y)$ is a polynomial of degree $L$ which satisfies
\begin{equation}
P^{(L)}(x,y)=P^{(L)}(y,x)=P^{(L)}(x,-2-x-y).
\end{equation}
Requiring these two conditions on a general polynomial of degree $L$ leaves $(L+2)(L+4)/8$ undetermined coefficients, which exactly agrees with the number of solutions found in the previous section.

 Before proceeding, let us mention that the supergravity solution in Mellin space can be written in the form \eqref{masp}, with
\begin{equation}
P^{(sugra)}(x,y)= \frac{16}{(x+1)(y+1)(1+x+y)}\,.
\end{equation}
In order to construct the most general solution we introduce the following set of variables
\begin{eqnarray}
s &=& x +2/3\,,\\
t &=& -4/3-x-y\,,\\
u&=&y+2/3\,.
\end{eqnarray}
These satisfy $s+t+u=0$ and as a consequence of the symmetries $P^{(L)}(s,t,u)$ should be a completely symmetric function in the three variables. Introducing
\begin{eqnarray}
\sigma_2 = s^2+t^2+u^2\,,\\
\sigma_3=s^3+t^3+u^3\,,
\end{eqnarray}
we can take our basis of solutions to be 
\begin{equation}
P_{p,q}(x,y)=\sigma_2^p \sigma_3^q\,,
\end{equation}
for non-negative integers $p,q$. These correspond to intermediate states up to spin $2(p+q)$. Finally, the supergravity solution can be written in these variables as
\begin{equation}
P^{(sugra)}(s,t,u)= -\frac{16}{(s+1/3)(t+1/3)(u+1/3)}\,.
\end{equation}

\subsection{Absence of other solutions}
The representation in Mellin space is also useful in order to discuss the existence of other solutions. Imagine we had an extra solution of the form
\begin{equation}
M^{extra}(x,y)= \frac{x^2 (1+x)^2 y^2 (1+y)^2}{(1-x-y)^2 (x+y)^2} f^{extra}(x,y),
\end{equation}
for a non-polynomial function $f^{extra}(x,y)$ satisfying the required symmetries. Let us consider the analytic structure of $f^{extra}(x,y)$ in the complex plane. An important feature of the Mellin space representation is that poles of $M(x,y)$ correspond to intermediate states. Having assumed the spectrum at large $N$, the structure of poles is already fixed. Then $f^{extra}(x,y)$ should have no poles. This means that $f^{extra}(x,y)$ is an entire function in the complex $x,y$ planes. If we assume polynomially bounded solutions, namely $\gamma_{n,\ell}$ grows at large $n$ at most as a polynomial, we require $f^{extra}(x,y)$ to be polynomially bounded as well\footnote{We assume that a non-polynomially bounded $M(x,y)$, consistent with crossing-symmetry, leads to a non-polynomially bounded $\gamma_{n,\ell}$. This was seen to be the case for all examples we have tried. The intuitive reason is that if $M(x,y)$ grows exponentially  inside an angular region, crossing symmetry would extend this region to other two regions via \eqref{sym} and \eqref{sym2} ({\it e.g.} the upper half plane in the complex $x-$plane is extended to $\Im (y)>0$ and $\Im (-2-x-y)>0$). The requirement of polynomially bounded Mellin amplitudes also arises if we require the CFT to have a dual description in terms of an effective field theory on $AdS$  \cite{Fitzpatrick:2012cg}.}. This implies that $f^{extra}(x,y)$ has to be a polynomial (since a polynomially bounded entire function is a polynomial). Therefore, such a function has to be a finite linear combination of the solutions we have already discussed. Let us mention that by independent arguments in \cite{Heemskerk:2009pn} it was shown, for non supersymmetric CFT, that all solutions are obtained as convergent sums of the bounded-spin solutions. There are two classes of solutions for which our assumptions do not hold and are unbounded in the spin. For one class, extra poles at large values $\Delta_{gap}$ are allowed. As we will argue, their contribution to higher spins is suppressed by powers of $\Delta_{gap}$. For the second class $\gamma_{n,\ell}$ is not polynomially bounded. As will be seen in the next section, such solutions would require a gap scale smaller than any positive power of $N$. Both classes correspond to sums of the solutions we have found, so in this sense our solutions are a complete set.


\section{Interpretation}
\label{Secion:Interpretation}
In the previous section we have obtained the general solution at order $1/N^2$ (and for large $g^2N$) consistent with crossing symmetry. It takes the final form
\begin{equation}
A(z,\bar z)= A^{(sugra)}(z,\bar z) + \sum_{p,q=0}^\infty \alpha_{p,q} A^{(p,q)}(z,\bar z) \,,
\end{equation}
where we have given explicit expressions for all solutions in Mellin space (and for the first few ones in space time). Note that the coefficient in front of $A^{(sugra)}(z,\bar z)$ is fixed, since $A^{(sugra)}(z,\bar z)$ cancels a contribution \eqref{firstorder} in $G^{(1)}_{short}(z,\bar z)$, which would violate our assumption for the spectrum. On the other hand, the solutions $A^{(p,q)}(z,\bar z)$ may have (in principle!) an arbitrary coefficient $\alpha_{p,q}$ in front. In this section we will analyze these solutions.


\subsection{Large \texorpdfstring{$n$}{} behavior}
For the discussion to follow it will be important to understand the contribution from each solution to the anomalous dimension $\gamma_{n,\ell}$ in the large $n$ limit. Let us start with the supergravity solution \eqref{sugraresult}, in the $n \gg \ell$ limit we obtain\footnote{The limit $\ell \gg n \gg 1$ will also be relevant below. In this case we obtain $\gamma^{sugra}_{n,\ell}= - 4 n^4/\ell^2$. \label{largel} }
\begin{equation}
\gamma^{sugra}_{n,\ell}   = -2 \frac{n^3}{\ell+1} +\ldots\,.
\end{equation}
We have given the explicit form of $A^{(p,q)}(z,\bar z)$ in Mellin space. It turns out that the large $n$ contribution to the anomalous dimensions from such solutions can be inferred from the large $x,y$ behaviour of the Mellin amplitude \cite{Fitzpatrick:2010zm,Fitzpatrick:2012cg}. More precisely
\begin{equation}
M(\rho \,x,\rho\, y) \sim \rho^s f(x,y)~~ \to  ~~ \gamma_{n}  \sim n^{2s+1}\,,
\end{equation}
at large $n$. One can indeed check that this gives the correct answer for the supergravity contribution. From the explicit form of $\sigma_2$ and $\sigma_3$ and the overall prefactor $M^{(0)}(x,y)$ we obtain
\begin{equation}
M^{(p,q)}(\rho \, x,\rho \, y) \sim \frac{x^4 y^4}{(x+y)^4}(x^2+ x y+y^2)^p (x y(x+y))^q\rho^{4+2p+3q}\,.
\end{equation}
Denoting by $\gamma^{(p,q)}_{n,\ell}$ the contribution to the anomalous dimension from $A^{(p,q)}(z,\bar z)$ we therefore obtain
\begin{equation}
\gamma^{(p,q)}_{n,\ell} \sim n^{4p+6q+9}\,.
\end{equation}
Note that even for the smallest $p,q$  the anomalous dimension grows quite fast with $n$. From this together with the relation $ a^{(1)}_{n,\ell} = \frac{1}{2}\frac{\partial}{\partial n} \left( a^{(0)}_{n,\ell}\gamma_{n,\ell} \right)$ we can obtain the behavior at large $n$ of the structure constants. The zeroth order structure constants behaves as
\begin{equation}
a^{(0)}_{n,\ell} \sim \frac{1+\ell}{2^\ell} \frac{n^2}{16^n}\,.
\end{equation}
Hence
\begin{equation}
a^{(1),sugra}_{n,\ell} \sim \frac{1}{2^\ell} \frac{n^5}{16^n},~~~~~a^{(p,q)}_{n} \sim - \frac{n^{4p+6q+11}}{16^n}\,.
\end{equation}


\subsection{Interpretation of our solutions}
The superconformal bootstrap equation was first proposed by \cite{Beem:2013qxa} and was used to find numerical bounds for the dimension of unprotected leading twist operators with low spin. In that paper it was observed that the numerical bounds for dimensions of operators with spin $l=0,2,4$ at large values of the central charge were consistent with the supergravity result
\begin{align}
\Delta_{0,0} &= 4 - \frac{16}{N^2}+\ldots\,,\\
\Delta_{0,2} &= 6 - \frac{4}{N^2}+\ldots\,,\\
\Delta_{0,4} &= 8 - \frac{48}{25 N^2}+\ldots\,.
\end{align}
In this section we argue that the solutions we have obtained, which are all of order $1/N^2$, are consistent with the results found by the numerical bootstrap.

As already mentioned in the introduction, the input we give to the conformal bootstrap equation is the spectrum of our putative CFT. This spectrum should be consistent with unitarity, namely, the dimensions should satisfy the unitarity bound and should lead to real OPE coefficients. Consider the spectrum taking into account only the supergravity solution $\gamma^{(sugra)}_{n,\ell}$. At large $n$ this behaves as
\begin{equation}
\Delta_{n,\ell} = 2n  -\frac{2}{N^2} \frac{n^3}{\ell+1} +\ldots\,.
\end{equation}
If we take $N$ very large but finite, at some large enough $n$ the spectrum will violate unitarity no matter how small $1/N$ is. Note that the sign in front of $\gamma^{(sugra)}_{n,\ell}$ is immaterial: if we had the opposite sign, the violation to unitarity would manifest in the sign of the square of the OPE coefficient. For the case at hand we see that we run into trouble when\footnote{Under mild assumptions and in order to preserve unitarity, the improved bound $\left|\gamma_{n,\ell}/N^2\right| < 4$ was derived in \cite{Fitzpatrick:2010zm} by using the optical theorem. }
\begin{equation}
\label{problems}
n \sim N\,.
\end{equation}
Even more importantly, note that at this point the subleading corrections $n^3/N^2$ are as large as the leading piece $2n$, and hence we cannot trust our perturbative solutions. This signals the fact that the spectrum should be modified at large $n$. More precisely we can trust $\gamma^{(sugra)}_{n,\ell}$ only below certain scale $\Delta_{gap}$. This is the scale which was assumed to be parametrically large when constructing the zero order spectrum. Around that scale new operators have to be included in the spectrum such that the exact spectrum is now consistent with unitarity:
\begin{equation}
\gamma^{(sugra)}_{n,\ell} \to \gamma^{(exact)}_{n,\ell}(\Delta_{gap}).
\end{equation}
Note that we are not assuming any particular dependence of $\Delta_{gap}$ with $N$. The relation \eqref{problems} implies an upper bound $\Delta_{gap} \lesssim N$, but the gap scale could be much smaller than that. In particular, $\Delta_{gap}$ could depend on other parameters to which the crossing relations are blind, such as the coupling constant. 

We can ask now, what is the expansion of $\gamma^{(exact)}_{n,\ell}(\Delta_{gap})$ in the limit $1 \ll n \ll \Delta_{gap}$. We expect the following behaviour 
\begin{equation}
\label{sugracomplete}
\gamma^{(exact)}_{n,\ell}(\Delta_{gap}) = -2 \frac{n^3}{\ell+1} + c_1 \frac{n^4}{\Delta_{gap}}+ c_2 \frac{n^5}{\Delta_{gap}^2}+\ldots\,,
\end{equation}
where we are first expanding in $1/\Delta_{gap}$ and then in large $n$. In order to understand this behavior it is convenient to consider the correlator in Mellin space and focus on the simplest example of exchange of a heavy operator of dimension $\Delta_{gap}$ (plus all its descendants) along the $s-$channel. This will produce a tower of poles in Mellin space, of the schematic form
\begin{equation}
\sum_{m} \frac{Res_{m}(y)}{2x+\Delta_{gap}+2m}
\end{equation}
corresponding to the sum over poles of the corresponding conformal block, see {\it e.g.} \cite{Fitzpatrick:2012cg}. As we take $\Delta_{gap}$ to be very large, the sum above localizes around $m \sim \Delta_{gap}^2$, effectively resulting in a pole at $x \sim \Delta_{gap}^2$. According to the previous discussion about the large $n$ behavior, we expect  then $\gamma^{(exact)}_{n,\ell}(\Delta_{gap}) \sim \gamma^{(sugra)}_{n,\ell} F(n/\Delta_{gap})$, plus subleading terms in $n$, leading to (\ref{sugracomplete}). In particular, $F(n/\Delta_{gap})$ has radius of convergence $n \sim \Delta_{gap}$.

Note that in  (\ref{sugracomplete}) as $n \sim \Delta_{gap}$ the higher orders in the expansion will start to contribute, so that problems with unitarity are potentially avoided. We argue that the full solution will be a a double expansion, in $1/N^2$ and $1/\Delta_{gap}$. Now comes a simple but important point: the extra terms in the expansion \eqref{sugracomplete} should also be consistent with crossing symmetry. Since in the previous section we have constructed, to order $1/N^2$, all solutions consistent with symmetry, the extra terms \eqref{sugracomplete} should be combinations of those. From the large $n$ behavior of $\gamma^{(p,q)}_{n,\ell}$ we conclude
\begin{equation}
\label{sugraours}
\gamma^{(exact)}_{n,\ell}(\Delta_{gap}) = \gamma^{(sugra)}_{n,\ell} + c_{0,0} \frac{\gamma^{(0,0)}_{n,\ell}}{\Delta_{gap}^6}+c_{1,0} \frac{\gamma^{(1,0)}_{n,\ell}}{\Delta_{gap}^{10}}+\ldots\,.
\end{equation}
The extra solutions we have found in particular capture the $1/\Delta_{gap}$ expansion of the exact completion of the supergravity solution. 

The necessity of an extra scale to render the spectrum consistent with unitarity was first discussed in \cite{Fitzpatrick:2010zm}, where it was motivated from the point of view of effective field theories in the $AdS$ bulk. From that point of view, the analogues of our extra solutions arise from non-renormalizable interactions in $AdS$ and are suppressed by powers of $\Delta_{gap}$, the powers being fixed by dimensional analysis. From a purely CFT point of view note that crossing symmetry allows the extra solutions with coefficients which are not suppressed
\begin{equation}
\label{final2}
\gamma(n,\ell) = \gamma^{sugra}_{n,\ell}+ \alpha_{0,0} \gamma^{(0,0)}_{n,\ell}+ \alpha_{1,0} \gamma^{(1,0)}_{n,\ell} +\ldots\,.
\end{equation}
The presence of extra solutions with non-suppressed overall coefficients will not be, in general, consistent with the numeric results quoted at the beginning of this section, unless precise linear inequalities are satisfied. Note that perturbative crossing symmetry alone is not sufficient to rule out such solutions, however, we would like to claim that such solutions are not present and the extra solutions appear always with suppressed overall coefficients. Below we present two compelling arguments for this claim, although we do not have a proof. 

First, note that (\ref{final2}) would imply that our solutions break down at smaller and smaller scales.  In order for the solutions not to break down, the simplest possibility is to assume $\Delta_{gap}$ is small enough, so that the spectrum changes, as described above \footnote{Another possibility is that one needs to consider the full, finite $N$, bootstrap equation. In that case none of our methods apply and we have nothing to say.}. More precisely, if \eqref{final2} includes $\gamma^{(p,q)}_{n,\ell}$ with coefficient $\alpha_{p,q} \sim 1$ we get the upper bound
\begin{equation}
\Delta_{gap} \lesssim N^{\frac{1}{4+2p+3q}}\,.
\end{equation}
For instance, including only the spin zero solution we would obtain $\Delta_{gap} \lesssim N^{1/4}$, while including also the next solution we would have $\Delta_{gap} \lesssim N^{1/6}$. From crossing symmetry considerations alone we are not able to find a lower bound for $\Delta_{gap}$. However, in the following we compare our solutions with known instanton contributions to the four point correlators. As we will see, this comparison suggests $\Delta_{gap} \sim N^{1/4}$, ruling out most of the extra solutions (or requiring suppressed overall coefficients).

\bigskip

{\noindent \bf Comparison with instanton solutions}

\vspace{0.15in}
\noindent Correlation functions in ${\cal N}=4$ SYM are known to receive non-perturbative instanton contributions \cite{Bianchi:1998nk,Dorey:1999pd}. In the large $N$ limit we expect the conformal bootstrap equation, and hence our treatment of it, to capture such solutions as well. The precise form of the instanton correction to the four-point function of protected operators considered in this paper was given in \cite{Arutyunov:2000im}. In our conventions, their result reads
\begin{equation}\label{instcorr}
{\cal G}(u,v)_{inst} = \frac{f(\tau)}{N^{7/2}} u^2 v^2 \bar{D}_{4444}(u,v),
\end{equation}
where $\tau$ is the complexified coupling constant. While $f(\tau)$ was computed in a semiclassical approximation (around the one-instanton background) the full solution, for all values of $\tau$, is expected to have this form  \cite{Bianchi:1998nk,Dorey:1999pd}. Hence it is valid to compare  this expression with our solutions. We see that \eqref{instcorr} has exactly the form $A^{(0,0)}(u,v)$! Furthermore the precise normalization is consistent with \eqref{sugraours}, and for a moderate coupling constant $g = \textrm{fixed}$, we obtain
\begin{equation}
\frac{1}{N^2}\frac{1}{\Delta_{gap}^6} \sim N^{-7/2} ~ \to ~ \Delta_{gap} \sim N^{1/4}\,,
\end{equation}
which coincides with the dimension of single trace operators in ${\cal N}=4$ SYM at large $N$ and $g = \textrm{fixed}$, see {\it e.g.} \cite{malda}. This strengthen our argument that all other solutions, besides the supergravity one, are suppressed. 

Before proceeding, note that according to the $AdS/CFT$ duality $\Delta_{gap}^2 \sim 1/\alpha'$. Therefore, in the dual picture the expansion \eqref{final2} corresponds to the expansion of the string theory result, as expected. 

\subsection{Connection to causality and UV completion}


{\noindent \bf UV completion and positivity constraints}


\vspace{0.15in}
\noindent Given a correlator in ${\cal N}=4$ SYM one can construct a corresponding scattering amplitude in flat space \cite{Okuda:2010ym,Penedones:2010ue}. The expression for the flat space amplitude follows directly from the Mellin expression for the correlator and for our particular case we obtain
\begin{equation}
\label{flatamplitude}
{\cal A}_{flat}(s,t,u) = -\frac{16}{s \, t \,u} +\sum_{p,q=0}  \alpha_{p,q} \sigma_2^p \sigma_3^q\,,
\end{equation}
where $\sigma_2=s^2+t^2+u^2$, $\sigma_3=s^3+t^3+u^3$ and $s+t+u=0$. We have suppressed an overall factor which depends on our precise conventions and $G_N \sim 1/N^2$. In \cite{Adams:2006sv} it was argued that there are positivity constraints on the $2 \to 2$ scattering in the forward direction $t \rightarrow 0$. More precisely, we can consider \eqref{flatamplitude} in the limit $t \to 0$. In this limit $\sigma_2 =2s^2$ and $\sigma_3 \to 0$ and we obtain
\begin{equation}
\label{flatamplitudeforward}
{\cal A}_{flat}(s,t,-s) = \frac{16}{s^2 \, t} + \alpha_{0,0}+ 2\alpha_{1,0}s^2+\ldots\,+ {\cal O}(t).
\end{equation}
According to \cite{Adams:2006sv}, the coefficients $ \alpha_{0,0}$, $\alpha_{1,0}$, etc., have to be positive, otherwise there is no hope to embed this amplitude into the amplitude for a UV complete theory, whose S-matrix has the usual analytic properties. 

What is the consequence of this fact for our discussion? Consider the first two extra solutions $A^{(0,0)}$  and  $A^{(1,0)}$. Their contribution to the leading-twist anomalous dimension for $\ell=0,2$ can be computed and we obtain
\begin{eqnarray}
\gamma_{0,0} &=& -16 - \frac{9}{7} \alpha_{0,0} -\frac{16}{7}\alpha_{1,0}\,,\\
\gamma_{0,2} &=& -4 -\frac{20}{11} \alpha_{1,0}\,,
\end{eqnarray}
where the contribution from supergravity has also been included. The positivity constraints $\alpha_{0,0}>0$, $\alpha_{1,0}>0$ lead to an upper bound for the anomalous dimension $\gamma_{0,0}$ and $\gamma_{0,2}$. This upper bound is consistent with the one found by the numerical conformal bootstrap! The situation is less straightforward if we include higher spin solutions. For instance, as $\sigma_3$ vanishes in the forward limit, the coefficient in front of most of our solutions is not constrained by these considerations. However, as we have seen above, for these solutions to be present we would need a quite small dimension gap. 

Finally, let us mention that this argument relies on the flat space limit of the $AdS/CFT$ duality and not solely on the CFT perspective. It should be possible to prove that a CFT with, lets say $\alpha_{0,0}<0$, has pathologies, along the lines of \cite{Adams:2006sv}.

\bigskip

{\noindent \bf Causality and large $n,\ell$ behavior}


\vspace{0.15in}
\noindent In \cite{malda} the graviton three-point coupling in weakly coupled theories of gravity was studied. For the case of asymptotically $AdS_D$ space the authors show that causality imposes non-trivial constraints on the anomalous dimensions $\gamma(n,\ell)$ of operators around large $N$. In the limit $\ell \gg n \gg 1$ they obtain
\begin{equation}
\label{cau1}
\gamma(n,\ell) \sim - \frac{n^{D-1}}{\ell^{D-3}}\,,
\end{equation}
where we have suppressed a factor of $G_N$ already implicit in our definition of $\gamma(n,\ell)$. This has been already derived using crossing arguments \cite{Alday:2007mf,Fitzpatrick:2012yx,Komargodski:2012ek}. Note that the supergravity result \eqref{sugraresult} exactly agrees with their result for $D=5$, see footnote \ref{largel}. In the opposite limit $n \gg \ell \gg 1$ (and $\frac{\ell}{n} > \frac{1}{\Delta_{gap}}$) they find
\begin{equation}
\label{cau2}
\gamma(n,\ell) \sim - n^2 \left(\frac{n}{\ell}\right)^{D-4}\,.
\end{equation}
Again, the supergravity result exactly agrees with this result. In the previous section we have obtained an infinite set of solutions to crossing equations. However, each of these solutions only contributes to a finite range of spins, hence, they will not contribute to the above limits\footnote{Note however that even though they are not forbidden, they are somehow disfavored, as they grow too fast in $n$.}. Solutions where the spin is unbounded are either suppressed or would require a very small gap scale, according to the discussion above. Hence, \eqref{cau1} and \eqref{cau2} follow from crossing symmetry. 


\section{Conclusions}
In this paper we have considered the four-point correlator of the stress tensor multiplet in ${\cal N}=4$ SCFT and have constructed all solutions consistent with crossing symmetry, as an expansion in $1/N^2$ and for fixed non-zero values of the coupling constant. In addition to the supergravity solution, necessary due to the structure of $F_{short}(u,v,c)$, we have found an infinite tower of solutions. Our solutions break down unless we introduce a scale $\Delta_{gap}$ \footnote{To be more precise, in order to construct our solutions we have assumed a parametrically large gap. Unitarity requires such a gap not to be too large.}. We argued that the extra tower of solutions is suppressed by powers of $\Delta_{gap}$. The full solution hence organizes as a double expansion in $1/c$ and $1/\Delta_{gap}$. Our solutions are valid to leading order in $1/c$ and to all orders in $1/\Delta_{gap}$. Comparison of our solutions with explicit instanton computations confirms our expectations and leads to $\Delta_{gap} \sim N^{1/4}$, which agrees with the known dimension of the operators neglected when constructing the zeroth order solution. This value of $\Delta_{gap}$ would imply most extra solutions would break down (unless suppressed). The basic reason is that if $\gamma_{n,\ell}$ for a given solution grows too fast with $n$, perturbation theory would break down, before $\Delta_{gap}$ enters into the game (which would be unexpected from a effective field theory point of view). Note that our solutions grow faster with $n$ than the solutions for a generic conformal field theory found in \cite{Heemskerk:2009pn}. This is due to supersymmetry and the shift $\Delta \to \Delta + 4$ in the definition of superconformal blocks. Actually, if we allow ourselves to use the improved bound $\left|\gamma_{n,\ell}/N^2\right| < 4$  derived in \cite{Fitzpatrick:2010zm}, then we can rule out all the extra solutions. This would explain why the extra solutions we have found, do not violate the upper bounds observed by \cite{Beem:2013qxa}. It would also be consistent with the fact that these bounds seem to work better as we increase the spin: the solutions entering at higher spin are suppressed by higher powers of $\Delta_{gap}$! Note however, that we have not proven the upper bounds observed by \cite{Beem:2013qxa}, as our solutions are valid only in the regime of large $g^2 N$. Besides, we have argued, but not proven, the fact that the extra solutions appear suppressed by the extra parameter $\Delta_{gap}$.

It would be interesting to see if $\Delta_{gap}$ can be determined entirely from the superconformal bootstrap equation, without any additional input. It is not clear to us if this can be the case.  In any case, note that with this single input our solutions reproduce much of the structure of the string theory result, at leading order in $1/N^2$ but to all orders in $\alpha'$ (or $1/\sqrt{\lambda}$).

We have also elucidated a connection between such upper bounds and positivity constraints arising from causality in flat space. This is not a purely CFT argument, and it would be interesting to show, following \cite{Adams:2006sv} that if a CFT correlator leads to flat space amplitudes with the wrong sign, then the CFT is pathologic. Note that these positivity constraints would lead to the correct upper bound even if the first solutions in the tower are not suppressed. This could be more relevant to applications where the growing with $n$ is slower. For instance, for standard four-dimensional conformal field theories and external operators with dimension two, the anomalous dimension for the first three solutions of \cite{Heemskerk:2009pn} grow like $n,n^5$ and $n^7$, while our first solution grows like $n^9$.

Finally, we have seen that certain relations for the anomalous dimension of double trace operators, derived from causality constraints for scattering in $AdS$ in \cite{malda} follow from crossing symmetry. It would be interesting to extend the positivity constraints of \cite{Adams:2006sv} to the case of $AdS$.

To summarize, trying to understand how the numeric conformal bootstrap for ${\cal N}=4$ SYM reproduces the supergravity result, we have learnt the following interesting lessons:
\begin{itemize}
\item The conformal bootstrap equation captures non-perturbative instanton solutions. As it should, since it is valid even non perturbatively, but here we are seeing this very explicitly.
\item With an additional input for $\Delta_{gap}$, the conformal bootstrap equation captures much of the structure of the full stringy result for the four-point correlator.  
\item The existence of upper bounds for the dimension of leading twist operators is related to positivity constraints arising from causality in flat space. 
\item Recent relations derived from causality  constraints for scattering in $AdS$ can be shown to follow from symmetry. 
\end{itemize}
These lessons indicate that for ${\cal N}=4$ SYM the conformal bootstrap equation knows not only about the supergravity result for anomalous dimensions but actually much more about the dual string theory.


\section*{Acknowledgements}
We are grateful to Zohar Komargodski and Juan Maldacena for enlightening discussions. This work was supported by ERC STG grant 306260. L.F.A. is
a Wolfson Royal Society Research Merit Award holder.


\appendix
\section{\texorpdfstring{$G_{short}(u,v,c)$}{} }
\label{App:Gshort}
In this appendix we summarize the form of the function\footnote{We thank the authors of \cite{Beem:2013qxa} for sharing the explicit form of this function with us.} $G_{short}(u,v,c)$ appearing in the conformal partial wave expansion performed in section \ref{analyticsolutionslarge}. Firstly, we make explicit its dependence on the central charge $c$
\begin{equation}
G_{short}(u,v,c)=G_{short}^{(0)}(u,v)+\frac{1}{c}G_{short}^{(1)}(u,v).
\end{equation}
Secondly, we organize both contributions separating various logarithmic terms 
\begin{equation}
G_{short}^{(i)}=R^{(i)}_1(z,\bar z)+R^{(i)}_2(z,\bar z)\log(1-z)+R^{(i)}_2(\bar z, z)\log(1-\bar z)+R^{(i)}_3(z,\bar z)\log(1-z)\log(1-\bar z).
\end{equation}
Then, all functions $R^{(i)}_j(z,\bar z)$ are rational and they take the following form
\begin{align}\nonumber
R^{(0)}_1&=\frac{4 \left(z^3 (\bar z-6) (\bar z-1)^2-8 z^2 (\bar z-1)^2 \bar z+z (\bar z (\bar z (13 \bar z-8)-46)+36)-6 \left(\bar z^3-6
   \bar z+4\right)\right)}{z \bar z}\,,\\\nonumber
 R^{(0)}_2&= \frac{8 (z-1)^2 \left(2 z \bar z^4-z \bar z^3+4 z \bar z^2-18 z \bar z+12 z-3 \bar z^4+18 \bar z^2-12 \bar z\right)}{z^2 \bar z (\bar z-z)}\,,\\\nonumber
R^{(0)}_3&=-\frac{96 (z-1)^2 (\bar z-1)^2}{z^2 \bar z^2}\,,\\\nonumber
R^{(1)}_1&=-\frac{4 (z-1) (\bar z-1) (17 z \bar z-18 z-18 \bar z+36)}{z \bar z}\,,\\\nonumber
R^{(1)}_2&=\frac{8 (z-1)^2 (\bar z-1) \left(4 z \bar z^2+9 z \bar z-18 z-9 \bar z^2+18 \bar z\right)}{z^2 \bar z (\bar z-z)}\,,\\
R^{(1)}_3&=-\frac{144 (z-1)^2 (\bar z-1)^2}{z^2 \bar z^2}\,.
\end{align}

\section{\texorpdfstring{$\bar D$}{}-functions}
\label{App:Dbar}
In this appendix we collect the definition and basic symmetries of the functions $\bar D_{\Delta_i}(u,v)$ we used in the main body of the paper. These functions enter in the computations of the Witten diagrams for the four-point function associated to the contact interactions in $AdS$. They are related to the function $D$ introduced in \cite{D'Hoker:1999pj}
\begin{equation}
D_{\Delta_i}(x_1,x_2,x_3,x_4)=\frac{\Gamma(\sfrac{1}{2}\sum_i\Delta_i-2)}{\prod_i\Gamma(\Delta_i)}\int_{0}^{\infty}\prod_i dt_i t_i^{\Delta_i-1}e^{-\sfrac{1}{2}\sum_{i,j}t_i t_j x_{ij}^2}
\end{equation}
in the following way
\begin{equation}
\bar D_{\Delta_i}(u,v)=\frac{2\prod_i\Gamma(\Delta_i)}{\Gamma(\sfrac{1}{2}\sum_i\Delta_i-2)}x_{13}^{2\Delta_1}x_{24}^{2\Delta_2}\left(\frac{x_{14}^2}{x_{13}^2x_{34}^2}\right)^{\sfrac{\Delta_1-\Delta_3}{2}}\left(\frac{x_{13}^2}{x_{14}^2 x_{34}^2}\right)^{\sfrac{\Delta_2-\Delta_4}{2}}D_{\Delta_i}(x_i).
\end{equation}
For the particular case $\Delta_i=1$, $\bar D_{\Delta_i}(u,v)$ reduces to the celebrated four-point scalar box integral, which in our conventions takes the form
\begin{equation}
\bar D_{1111}(u,v)=\frac{1}{z-\bar z}\left(2\,\mbox{Li}_2(z)-2\,\mbox{Li}_2(\bar z)+\log(z\,\bar z)\log\frac{1-z}{1-\bar  z}\right).
\end{equation}
For computational purposes it is convenient to construct all other functions by acting on $\bar D_{1111}$ with differential operators introduced in \cite{Arutyunov:2002fh}.

In order to prove symmetries of all our solutions in section \ref{analyticsolutionslarge} we used the following symmetries of $\bar D$-functions
\begin{align}\nonumber
\bar D_{\Delta_1\Delta_2\Delta_3\Delta_4}(u,v)&=\bar D_{\Delta_3\Delta_2\Delta_1\Delta_4}(v,u)\\
&=v^{\Delta_4-\sfrac{1}{2}\sum_i\Delta_i}\bar D_{\Delta_2\Delta_1\Delta_3\Delta_4}(\sfrac{u}{v},\sfrac{1}{v}).
\end{align}


\bibliographystyle{nb}
\bibliography{bibliography}

\begin{thebibliography}{10}
\ifx\href\asklfhas\newcommand{\href}[2]{#2}\fi
\ifx\arxivref\asklfhas\newcommand{\arxivref}[2]{\href{http://arxiv.org/abs/#1}{#2}}\fi
\ifx\doiref\asklfhas\newcommand{\doiref}[2]{\href{http://dx.doi.org/#1}{#2}}\fi
\raggedright
\small
\parskip 0pt

\bibitem{Beem:2013qxa}
C.~Beem, L.~Rastelli and B.~C.~van~Rees,
\textit{``{The N=4 Superconformal Bootstrap}''},
\textsf{\doiref{10.1103/PhysRevLett.111.071601}{Phys.Rev.Lett.~111,~071601~(2013)}},
\texttt{\arxivref{1304.1803}{arxiv:1304.1803}}.

\bibitem{Nirschl:2004pa}
M.~Nirschl and H.~Osborn,
\textit{``{Superconformal Ward identities and their solution}''},
\textsf{\doiref{10.1016/j.nuclphysb.2005.01.013}{Nucl.Phys.~B711,~409~(2005)}},
\texttt{\arxivref{hep-th/0407060}{hep-th/0407060}}.

\bibitem{Rattazzi:2008pe}
R.~Rattazzi, V.~S.~Rychkov, E.~Tonni and A.~Vichi,
\textit{``{Bounding scalar operator dimensions in 4D CFT}''},
\textsf{\doiref{10.1088/1126-6708/2008/12/031}{JHEP~0812,~031~(2008)}},
\texttt{\arxivref{0807.0004}{arxiv:0807.0004}}.

\bibitem{D'Hoker:1999jp}
E.~D'Hoker, S.~D.~Mathur, A.~Matusis and L.~Rastelli,
\textit{``{The Operator product expansion of N=4 SYM and the 4 point functions
  of supergravity}''},
\textsf{\doiref{10.1016/S0550-3213(00)00523-X}{Nucl.Phys.~B589,~38~(2000)}},
\texttt{\arxivref{hep-th/9911222}{hep-th/9911222}}.

\bibitem{Arutyunov:2000ku}
G.~Arutyunov, S.~Frolov and A.~C.~Petkou,
\textit{``{Operator product expansion of the lowest weight CPOs in N=4 SYM(4)
  at strong coupling}''},
\textsf{\doiref{10.1016/S0550-3213(00)00439-9}{Nucl.Phys.~B586,~547~(2000)}},
\texttt{\arxivref{hep-th/0005182}{hep-th/0005182}}.

\bibitem{Dolan:2001tt}
F.~Dolan and H.~Osborn,
\textit{``{Superconformal symmetry, correlation functions and the operator
  product expansion}''},
\textsf{\doiref{10.1016/S0550-3213(02)00096-2}{Nucl.Phys.~B629,~3~(2002)}},
\texttt{\arxivref{hep-th/0112251}{hep-th/0112251}}.

\bibitem{Heemskerk:2009pn}
I.~Heemskerk, J.~Penedones, J.~Polchinski and J.~Sully,
\textit{``{Holography from Conformal Field Theory}''},
\textsf{\doiref{10.1088/1126-6708/2009/10/079}{JHEP~0910,~079~(2009)}},
\texttt{\arxivref{0907.0151}{arxiv:0907.0151}}.

\bibitem{Fitzpatrick:2012yx}
A.~L.~Fitzpatrick, J.~Kaplan, D.~Poland and D.~Simmons-Duffin,
\textit{``{The Analytic Bootstrap and AdS Superhorizon Locality}''},
\textsf{\doiref{10.1007/JHEP12(2013)004}{JHEP~1312,~004~(2013)}},
\texttt{\arxivref{1212.3616}{arxiv:1212.3616}}.

\bibitem{Komargodski:2012ek}
Z.~Komargodski and A.~Zhiboedov,
\textit{``{Convexity and Liberation at Large Spin}''},
\textsf{\doiref{10.1007/JHEP11(2013)140}{JHEP~1311,~140~(2013)}},
\texttt{\arxivref{1212.4103}{arxiv:1212.4103}}.

\bibitem{Alday:2013cwa}
L.~F.~Alday and A.~Bissi,
\textit{``{Higher-spin correlators}''},
\textsf{\doiref{10.1007/JHEP10(2013)202}{JHEP~1310,~202~(2013)}},
\texttt{\arxivref{1305.4604}{arxiv:1305.4604}}.

\bibitem{Fitzpatrick:2010zm}
A.~L.~Fitzpatrick, E.~Katz, D.~Poland and D.~Simmons-Duffin,
\textit{``{Effective Conformal Theory and the Flat-Space Limit of AdS}''},
\textsf{\doiref{10.1007/JHEP07(2011)023}{JHEP~1107,~023~(2011)}},
\texttt{\arxivref{1007.2412}{arxiv:1007.2412}}.

\bibitem{malda}
X.~O.~Camanho, J.~D.~Edelstein, J.~Maldacena and A.~Zhiboedov,
\textit{``{Causality Constraints on Corrections to the Graviton Three-Point
  Coupling}''},
\texttt{\arxivref{1407.5597}{arxiv:1407.5597}}.

\bibitem{Dolan:2006ec}
F.~Dolan, M.~Nirschl and H.~Osborn,
\textit{``{Conjectures for large N superconformal N=4 chiral primary four point
  functions}''},
\textsf{\doiref{10.1016/j.nuclphysb.2006.05.009}{Nucl.Phys.~B749,~109~(2006)}},
\texttt{\arxivref{hep-th/0601148}{hep-th/0601148}}.

\bibitem{Fitzpatrick:2011dm}
A.~L.~Fitzpatrick and J.~Kaplan,
\textit{``{Unitarity and the Holographic S-Matrix}''},
\textsf{\doiref{10.1007/JHEP10(2012)032}{JHEP~1210,~032~(2012)}},
\texttt{\arxivref{1112.4845}{arxiv:1112.4845}}.

\bibitem{Penedones:2010ue}
J.~Penedones,
\textit{``{Writing CFT correlation functions as AdS scattering amplitudes}''},
\textsf{\doiref{10.1007/JHEP03(2011)025}{JHEP~1103,~025~(2011)}},
\texttt{\arxivref{1011.1485}{arxiv:1011.1485}}.

\bibitem{Fitzpatrick:2011ia}
A.~L.~Fitzpatrick, J.~Kaplan, J.~Penedones, S.~Raju and B.~C.~van~Rees,
\textit{``{A Natural Language for AdS/CFT Correlators}''},
\textsf{\doiref{10.1007/JHEP11(2011)095}{JHEP~1111,~095~(2011)}},
\texttt{\arxivref{1107.1499}{arxiv:1107.1499}}.

\bibitem{Paulos:2011ie}
M.~F.~Paulos,
\textit{``{Towards Feynman rules for Mellin amplitudes}''},
\textsf{\doiref{10.1007/JHEP10(2011)074}{JHEP~1110,~074~(2011)}},
\texttt{\arxivref{1107.1504}{arxiv:1107.1504}}.

\bibitem{Mack:2009mi}
G.~Mack,
\textit{``{D-independent representation of Conformal Field Theories in D
  dimensions via transformation to auxiliary Dual Resonance Models. Scalar
  amplitudes}''},
\texttt{\arxivref{0907.2407}{arxiv:0907.2407}}.

\bibitem{Fitzpatrick:2011hu}
A.~L.~Fitzpatrick and J.~Kaplan,
\textit{``{Analyticity and the Holographic S-Matrix}''},
\textsf{\doiref{10.1007/JHEP10(2012)127}{JHEP~1210,~127~(2012)}},
\texttt{\arxivref{1111.6972}{arxiv:1111.6972}}.

\bibitem{Fitzpatrick:2012cg}
A.~L.~Fitzpatrick and J.~Kaplan,
\textit{``{AdS Field Theory from Conformal Field Theory}''},
\textsf{\doiref{10.1007/JHEP02(2013)054}{JHEP~1302,~054~(2013)}},
\texttt{\arxivref{1208.0337}{arxiv:1208.0337}}.

\bibitem{Bianchi:1998nk}
M.~Bianchi, M.~B.~Green, S.~Kovacs and G.~Rossi,
\textit{``{Instantons in supersymmetric Yang-Mills and D instantons in IIB
  superstring theory}''},
\textsf{\doiref{10.1088/1126-6708/1998/08/013}{JHEP~9808,~013~(1998)}},
\texttt{\arxivref{hep-th/9807033}{hep-th/9807033}}.

\bibitem{Dorey:1999pd}
N.~Dorey, T.~J.~Hollowood, V.~V.~Khoze, M.~P.~Mattis and S.~Vandoren,
\textit{``{Multi-instanton calculus and the AdS / CFT correspondence in N=4
  superconformal field theory}''},
\textsf{\doiref{10.1016/S0550-3213(99)00193-5}{Nucl.Phys.~B552,~88~(1999)}},
\texttt{\arxivref{hep-th/9901128}{hep-th/9901128}}.

\bibitem{Arutyunov:2000im}
G.~Arutyunov, S.~Frolov and A.~Petkou,
\textit{``{Perturbative and instanton corrections to the OPE of CPOs in N=4
  SYM(4)}''},
\textsf{\doiref{10.1016/S0550-3213(01)00118-3}{Nucl.Phys.~B602,~238~(2001)}},
\texttt{\arxivref{hep-th/0010137}{hep-th/0010137}}.

\bibitem{Okuda:2010ym}
T.~Okuda and J.~Penedones,
\textit{``{String scattering in flat space and a scaling limit of Yang-Mills
  correlators}''},
\textsf{\doiref{10.1103/PhysRevD.83.086001}{Phys.Rev.~D83,~086001~(2011)}},
\texttt{\arxivref{1002.2641}{arxiv:1002.2641}}.

\bibitem{Adams:2006sv}
A.~Adams, N.~Arkani-Hamed, S.~Dubovsky, A.~Nicolis and R.~Rattazzi,
\textit{``{Causality, analyticity and an IR obstruction to UV completion}''},
\textsf{\doiref{10.1088/1126-6708/2006/10/014}{JHEP~0610,~014~(2006)}},
\texttt{\arxivref{hep-th/0602178}{hep-th/0602178}}.

\bibitem{Alday:2007mf}
L.~F.~Alday and J.~M.~Maldacena,
\textit{``{Comments on operators with large spin}''},
\textsf{\doiref{10.1088/1126-6708/2007/11/019}{JHEP~0711,~019~(2007)}},
\texttt{\arxivref{0708.0672}{arxiv:0708.0672}}.

\bibitem{D'Hoker:1999pj}
E.~D'Hoker, D.~Z.~Freedman, S.~D.~Mathur, A.~Matusis and L.~Rastelli,
\textit{``{Graviton exchange and complete four point functions in the AdS / CFT
  correspondence}''},
\textsf{\doiref{10.1016/S0550-3213(99)00525-8}{Nucl.Phys.~B562,~353~(1999)}},
\texttt{\arxivref{hep-th/9903196}{hep-th/9903196}}.

\bibitem{Arutyunov:2002fh}
G.~Arutyunov, F.~Dolan, H.~Osborn and E.~Sokatchev,
\textit{``{Correlation functions and massive Kaluza-Klein modes in the AdS /
  CFT correspondence}''},
\textsf{\doiref{10.1016/S0550-3213(03)00448-6}{Nucl.Phys.~B665,~273~(2003)}},
\texttt{\arxivref{hep-th/0212116}{hep-th/0212116}}.

\end{thebibliography}
\end{document}